\documentclass[reprint,nofootinbib]{revtex4}
\usepackage{graphicx}
\begin{document}

\title{Entropic control of particle sizes during viral self-assembly}

\author{M Castelnovo$^1$, D Muriaux$^2$, and C Faivre-Moskalenko$^1$}

\address{$^1$ Laboratoire de Physique, Ecole Normale Sup\'erieure de Lyon, Universit\'e de Lyon and CNRS, 46 All\'ee d'Italie, 69364 LYON Cedex 07, FRANCE}
\address{$^2$ Centre d'\'etudes d'agents Pathog\`enes et Biotechnologies pour la Sant\'e, UMR5236, 1919 Route de Mende, 34293 MONTPELLIER, FRANCE}
\begin{abstract}
Morphologic diversity is observed across all families of viruses. Yet these supra-molecular assemblies are produced most of the time in a spontaneous way through complex molecular self-assembly scenarios. The modeling of these phenomena remains a challenging problem within the emerging field of \textit{Physical Virology}. We present in this work a theoretical analysis aiming at highlighting the particular role of configuration entropy in the control of viral particle size distribution. Specializing this model to retroviruses like HIV-1, we predict a new mechanism of entropic control of both RNA uptake into the viral particle, and of the particle's size distribution. Evidence of this peculiar behavior has been recently reported experimentally.
\end{abstract}
\maketitle

\section{Introduction}
Viruses rely mainly on molecular self-assembly to perpetuate their life cycle. Spontaneous self-assembly of molecules is indeed a powerful and yet passive way of structuring solutions of molecules at an intermediate length scale between the nanoscopic scale of the molecule itself and the microscopic scale of cells. Viruses are composed of proteins, nucleic acids, and eventually lipids in the case of enveloped viruses. All these components have to be orchestrated in order to produce the regular morphologies observed across different viral families through self-assembly \cite{lyonnais2012}. The precise modeling of these phenomena is generally challenging. In particular the question of the necessary regulation or control of self-assembly remains largely open for viruses with complex life cycles.

In the case of non-enveloped virus, the genome of the virus is protected by a protein shell called a capsid. The proteins are arranged according to icosahedral symmetry \cite{zandi2004}. As a consequence, this symmetry imposes some restrictions on the size distribution of viral particles to be self-assembled. Spontaneous curvature of protein layers has been put forward recently as another plausible mechanism of the control of the size polydispersity using continuous models of the capsid \cite{siber2009}. In the present work, we would like to emphasize the particular role of the genome in the regulation of particle size distribution. This is especially important in the case of viruses which have multipartite genomes. Indeed, for such viruses the configuration entropy of all the molecules constituting the virus is of utmost relevance and its balance with enthalpic contribution to the self-assembly leads to the appearance of specific phenomena like the entropic control of particle size distribution discussed in the present work. Zandi and Van der Schoot discussed recently within a similar formalism the interplay between electrostatic forces driving co-assembly of proteins and RNA and their relative stoichiometry \cite{zandi2009}. Following their work, we extend the modeling of viral self-assembly in order to describe the competition between different particles sizes and different RNA content. Our analysis leads to the following identification of several roles of entropic origin for the genome: \textit{(i)} the genome facilitates the viral self-assembly of proteins by lowering the onset of particle formation and by increasing the effective free energy gain per protein upon particle formation; \textit{(ii)} viral RNAs are preferentially co-packaged based on entropic considerations within viral particles in the mixtures of viral and cellular RNAs; \textit{(iii)} the uptake of viral genome produces a shift of particle size distribution towards smaller particles and a reduced polydispersity. 

This paper is organized as follows. In the first part, we present the classical thermodynamic framework to describe micellization phenomena in the case of a monodisperse protein self-assembly. The entropic role of the genome on the self-assembly is then investigated. The second part describes the influence of viral and cellular RNA uptake on the self-assembly process. The entropic uptake of viral RNA and the entropic control of size polydispersity found using the model is discussed with respect to recent experiments performed on HIV-1.
\section{Viral self-assembly and the entropic role of the genome}
\subsection{Classical description of pure protein self-assembly using micellization thermodynamics}
We consider in this section identical proteins that have a spontaneous tendency to self-assemble into a set of different aggregates. Knowing the input concentration of proteins $\phi_o$, we investigate the equilibrium partitioning of proteins into the different aggregates. Each aggregates is made of $p$ proteins, and the equilibrium concentration of these aggregates is written as $c_p$. The gain in free energy for the formation of one aggregate of size $p$ is $kTF_p$, where $k$ is the Boltzmann constant and $T$ the temperature of the system. The reference free energy of a single protein is $kTF_1$. The Gibbs free energy of the solution of proteins is written as 
\begin{equation}
\frac{G}{VkT}=c_1 (\ln \left(c_1 v_0\right)-1+F_1)+\sum_{p=2}^{\infty}c_p (\ln \left(c_p v_0\right)-1+F_p)
\end{equation}
where $V$ is the volume of the solution, and $v_0$ is a reference volume, interpreted as the cell volume used to compute the configuration entropy. 
For each aggregate type, there is a translational entropy term $kTVc_p (\ln \left(c_p v_0\right)-1)$ and an energetic gain term for the formation of  aggregate  $kTVc_pF_p$. As it is described below, this is the balance between these entropic and enthalpic contributions that sets the precise size distribution.

This Gibbs free energy assumes implicitly that long-range interactions between aggregates are negligible. At equilibrium, the size distribution $c_p$ minimizes the Gibbs free energy with the global constraint of mass conservation
\begin{equation}
\phi_0=c_1+\sum_{p=2}^{\infty}pc_p
\end{equation}
This can be taken into account by the use of a Lagrange multiplier $\mu$ that is interpreted as the chemical potential of individual proteins. The equilibrium conditions are written as
\begin{eqnarray}
c_p v_0& = & \left(c_1v_0\right)^pe^{-(F_p-pF_1)} \label{mass action}\\
\phi_0 v_0& = & c_1v_0+\sum_{p=2}^{\infty}p(c_1v_0)^pe^{-(F_p-pF_1)}	
\end{eqnarray}
The first equation is simply the law of mass action for the aggregate of size $p$. Using the notation $\Delta G_p\equiv F_p-pF_1\equiv pg_p$, one can find the equilibrium partition of proteins among the 
different aggregates by solving the following  non-linear equation in $c_1$
\begin{equation}
\phi_0 v_0= c_1v_0+\sum_{p=2}^{\infty}p(c_1v_0)^pe^{-pg_p}
\end{equation}
 and by plugging the solution into the law of mass action Eq.\ref{mass action}.
In order to address the question of dominance of a given population of particles with respect to another one, we restrict the model to a bimodal size distribution: the product of the self-assembly is either a small particle with $p_1$  proteins or a large particle with $p_2$ proteins. The equilibrium concentration of un-aggregated proteins $c_{1}$ is now given by
\begin{equation}
\phi_0 v_0 = c_1v_0+p_1(c_1v_0)^{p_1}e^{-p_1g_1}+p_2(c_1v_0)^{p_2}e^{-p_2g_2}
\end{equation}

\begin{figure}[htbp]
\begin{center}
\includegraphics[scale=0.35]{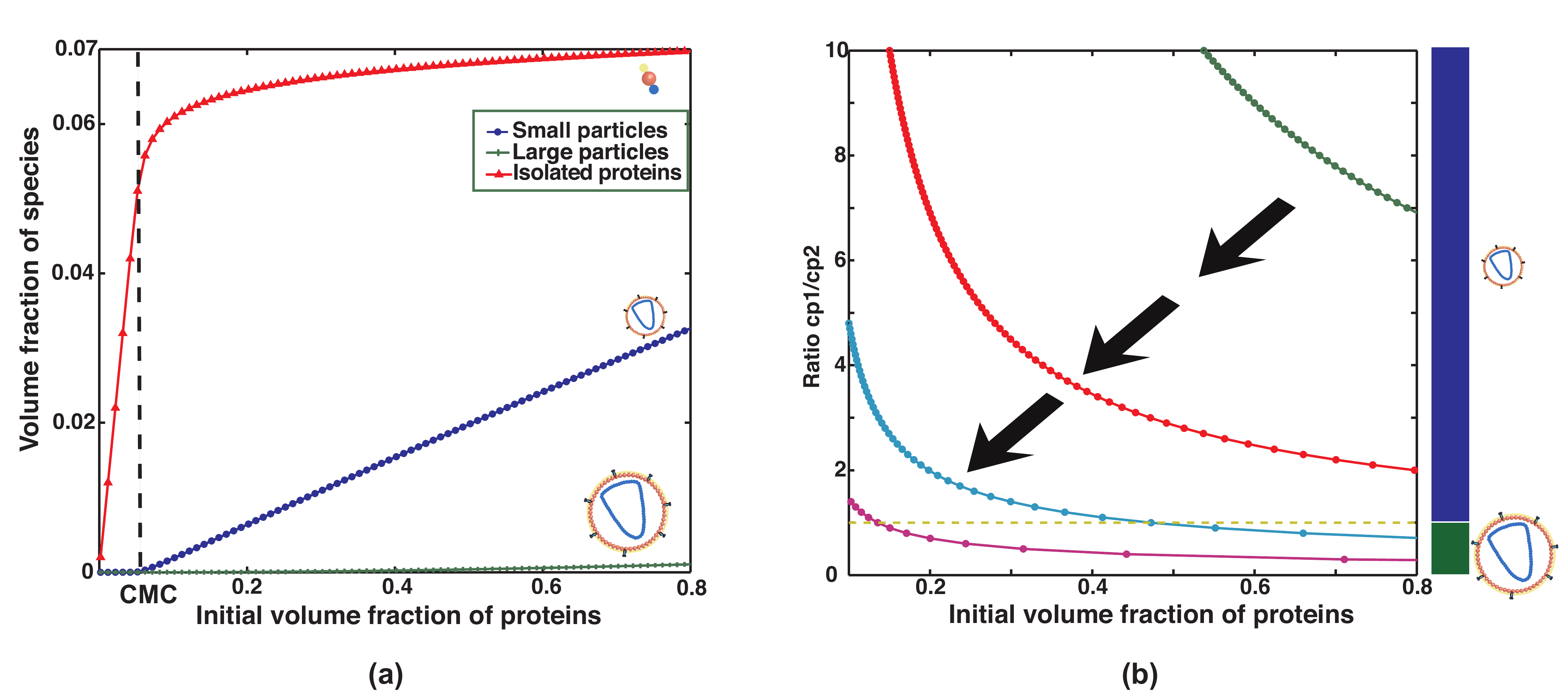}
\caption{Protein self-assembly into bimodal capsids. \textit{(a)} Protein titration in the absence of enthalpic selection ($g_1=g_2$). The onset of aggregation, known as the critical micellar concentration  (cmc) is shown by the black dashed line. The small particles are more numerous than large particles once the aggregation begins. Parameters are $p_1=21,p_2=42,g_1=g_2=-2.5,vo=2nm^3$. \textit{(b)} Ratio $c_{p_1}/c_{p_2}$ as function of initial concentration. The curves correspond to increasing difference $g_1-g_2>0$ (favoring larger particles) between enthalpic gain following the large black arrows. The orange dashed line represents the limit between small particle dominance (blue rectangle) and large particle dominance (green rectangle). The entropic selection reduces as the enthalpic gain is increased. Parameters are identical to \textit{(a)}, except for $g_1=-2.5$, $g_2=-2.54$ (green), $g_2=-2.58$ (red), $g_2=-2.62$ (cyan), $g_2=-2.66$ (pink). }
\label{figure1}
\end{center}
\end{figure}

The figure \ref{figure1}a shows the numerical resolution of the previous equation for a representative  set of parameters mimicking a protein titration experiment. This set of parameters was chosen because the representation of the results are in this case particularly clear. We checked however that the results described below are not strongly dependent on the precise choise of the parameters. In the particular case where the free energy gains per protein for small particles $g_1$ and for large particles $g_2$ are equal, any imbalance between population of small and large particles is directly attributed to purely entropic effect. For the sake of convenience, this scenario is called the "non-selective enthalpy"  (NSE) scenario.

For low initial concentration of proteins, the entropy of individual proteins cannot be balanced by free energy gain per protein $g_i$, and no self-assembly occurs. Once the so-called critical micellar concentration (CMC) is reached, protein association starts\footnote{Beyond this threshold, both isolated proteins, small and large particles concentrations change their behavior as more proteins are added to the solution}. This threshold is roughly estimated by
\begin{equation}
\label{CMC}
\phi_0 v_0\simeq \left(\frac{e^{p_1 g_1}}{p_1}\right)^{\frac{1}{p_1-1}} 
\end{equation}
Once the aggregation sets in, it is observed that there are more smaller particles than larger particles. This is understood as a purely entropic effect, as it was already mentioned earlier: indeed, a larger number of smaller particles can be formed at fixed concentration of proteins. 

It is possible to relax the NSE model by choosing distinct free energy gains per protein\footnote{In order to obtain spontaneous self-assembly, free energy gain per protein must be negative $g_1,g_2<0$} $g_1>g_2$. In this case, the enthalpy contribution to self-assembly tends to favor larger particles, and therefore it will counterbalance the entropic selection mechanism illustrated in figure \ref{figure1}a. After little algebra, it is possible to find an exact relationship between the concentration of initial protein $\phi_0$ and the ratio between the equilibrium value of the number of small and large particles $\alpha=c_{p1}/c_{p2}$
\begin{equation}
\phi_0 v_0=\left(\alpha e^{(p_1g_1-p_2g_2)}\right)^{\frac{1}{p_1-p_2}}+(p_1+\frac{p_2}{\alpha})\left(\alpha e^{(p_1g_1-p_2g_2)}\right)^{\frac{p_1}{p_1-p_2}}e^{-p_1g_1}
\end{equation} 

The solution to this equation $\alpha=f(\phi_0)$ is shown in figure \ref{figure1}b for different values of $g_1-g_2$. A progressive loss of entropic selection compared to the enthalpic selection is observed. As a consequence the entropic selection of small particles is therefore subjected to an assumption of NSE scenario and might be observed only under weak enthalpic size selectivity.

\subsection{Entropic role of monodisperse RNA upon protein self-assembly}
The previous calculation is useful at illustrating the role of entropy in the partitioning of proteins among different particles. However, many viruses require the presence of their genome in order to initiate or complete their assembly. Without specifying the precise structure of the capsid with its inner genome (single-stranded RNA in most cases), it is possible to generalize the previous approach in order to predict the influence of a monodisperse genome on the viral particle self-assembly. This generalization has been partly done in reference \cite{zandi2009}, and therefore the results of this section are similar to those obtained earlier.

The first important feature to be incorporated into this generalized model is the stoichiometry of viral particles. Indeed, several works have recently pointed out a linear relation of electrostatic origin between the number of proteins in the capsid and the total number of nucleotides in the genome\cite{zandi2009,belyi2006,ting2011,hu2008}. Following these works, we will assume that for each particle made of $p_i$ proteins, there are $m_i$ RNA molecules such that $p_i=Km_i$, where $K$ is a constant depending in particular on the length of RNA. Assuming that the initial concentrations of proteins and RNAs are respectively $\phi_0$ and $\phi_r$, the equilibrium equations for self-assembly are easily written as
\begin{eqnarray}
\phi_0 & = & c_0+\sum_{i} p_ic_{p_i}  \label{phioR}\\
\phi_{r} & = & c_{r}+\sum_{i}m_ic_{p_i}  \label{phirR}\\
c_{p_i} & = & c_0^{p_i}c_{r}^{m_i}e^{-p_ig_i} v_0^{p_i+m_i-1} \label{cpR}\\
\end{eqnarray} 
where $c_r$ is the concentration of free RNAs.
In order to highlight the role of genome in the self-assembly process, we will restrict the product of self-assembly to two distinct sizes of capsid made of $p_1$ and $p_2$ proteins. Furthermore, each of these particles $p_{1,2}$ have the possibility to contain either no RNA or $m_{1,2}$ RNAs. Therefore there are four types of particles within this model (two particle sizes and two RNA contents each), and this new feature goes beyond the two particle treatment performed in reference \cite{zandi2009}. This particular configuration allows to show two main features of RNA presence during the self-assembly. The first one is the enhancement of  the entropic selection of smaller particles containing RNA even if RNA uptake is done without extra free energy gain. The second one is the lowering of the CMC for particle assembly.

\begin{figure}[htbp]
\begin{center}
\includegraphics[scale=0.35]{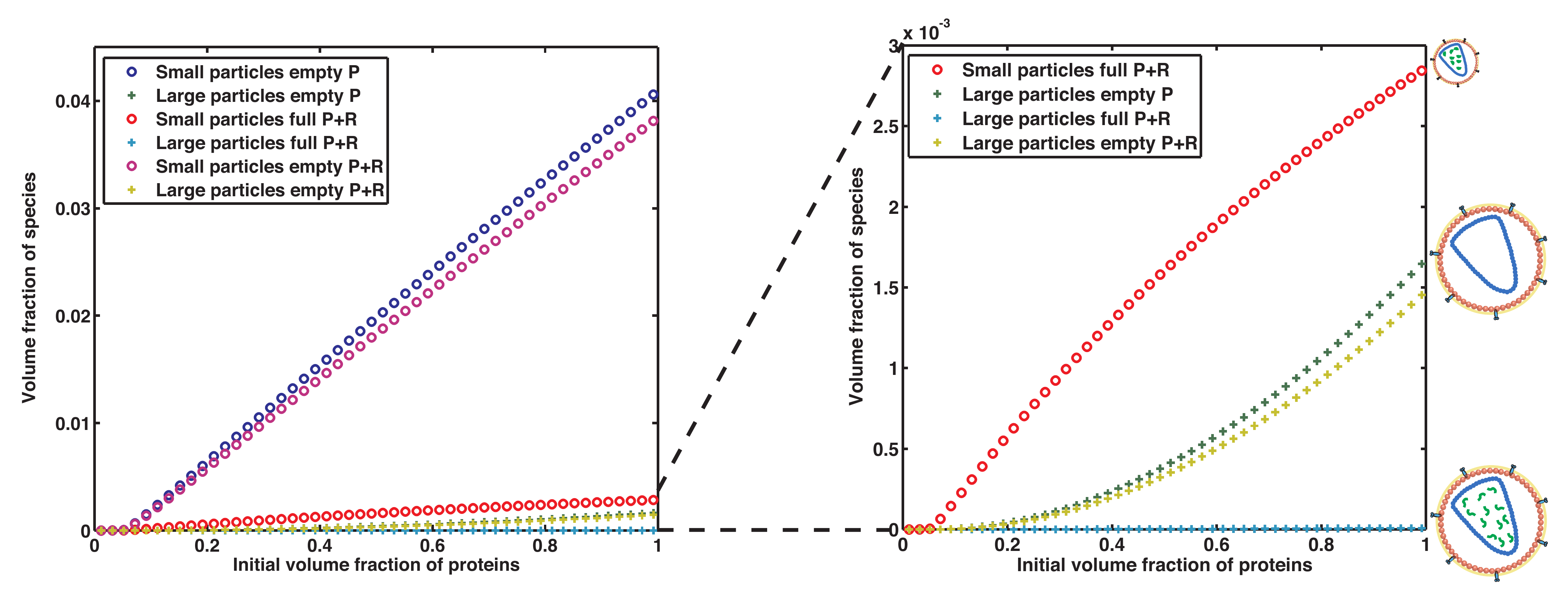}
\caption{Enhancement of entropic selection by the presence of RNA. \textit{(a)} Concentration of particles in the case of pure protein self-assembly (P) and in the presence of RNA (P+R) during protein titration. In the latter case, four types are particles are formed. No extra gain in free energy due to the presence of RNA was assumed. Parameters are: $p_1=21,p_2=42,g_1^{(P)}=g_1^{(P+R)}=g_2^{(P)}=g_2^{(P+R)}=-2.5,v0=2nm^3,\phi_rv_0=0.8$. \textit{(b)} Selection of previous graph showing that the full small particles are more numerous than larger particles in all cases (full or empty in the assembly in the presence of RNA, and empty in the pure protein assembly). }
\label{figure2}
\end{center}
\end{figure}

The first feature is illustrated in figure \ref{figure2}. For the sake of notation clarity, we define the free energy gain per protein in the presence or in the absence of RNA respectively by $g_i^{(P+R)}$ and $g_i^{(P)}$. In the case where RNA does not bring extra gain in free energy upon their uptake in viral particle, we have  $g_i^{(P+R)}=g_i^{(P)}$, and the results of figure \ref{figure2} shows that small particles containing RNA are more numerous than larger particles regardless of their RNA content. This shows that the presence of the genome in the solution is a key factor affecting the relative populations of particles. 

Rewriting the equilibrium equations, this can be understood as an effective increase in free energy gain per protein. Indeed the equations \ref{phioR},\ref{phirR} and \ref{cpR} applied to the four types particles are written as:
 \begin{eqnarray}
 \phi_0 & = & c_0 +p_1c_0^{p_1}e^{-p_1g_{1eff}}+p_2c_0^{p_2}e^{-p_2g_{2eff}} \\
 g_{ieff} & = & g_i^{(0)}-\frac{\ln \left(1+c_r^{m_i}e^{-p_i\delta g_i}v_0^{m_i} \right)}{p_i}
 \end{eqnarray}
where $g_i^{(0)}$ is the free energy gain per protein for pure protein self-assembly into particle of size $p_i$, $\delta g_i$ is the extra free energy gain per protein brought by the presence of RNA. This last term contains in particular both the contribution of RNA entropy within the particle and the specificity of RNA-protein interactions. Their contributions to the size selection phenomena discussed with our formalism are expected to produce similar effects. The relative contribution from RNA entropy and RNA-protein interactions is however expected to be more model-dependent, and goes therefore beyond the scope of the present work. This last equation shows that even without intrinsic extra free energy gain $\delta g_i=0$, the entropy of RNA allows to effectively increase the free energy gain, $g_{ieff}$ becoming more negative.

\begin{figure}[htbp]
\begin{center}
\includegraphics[scale=0.35]{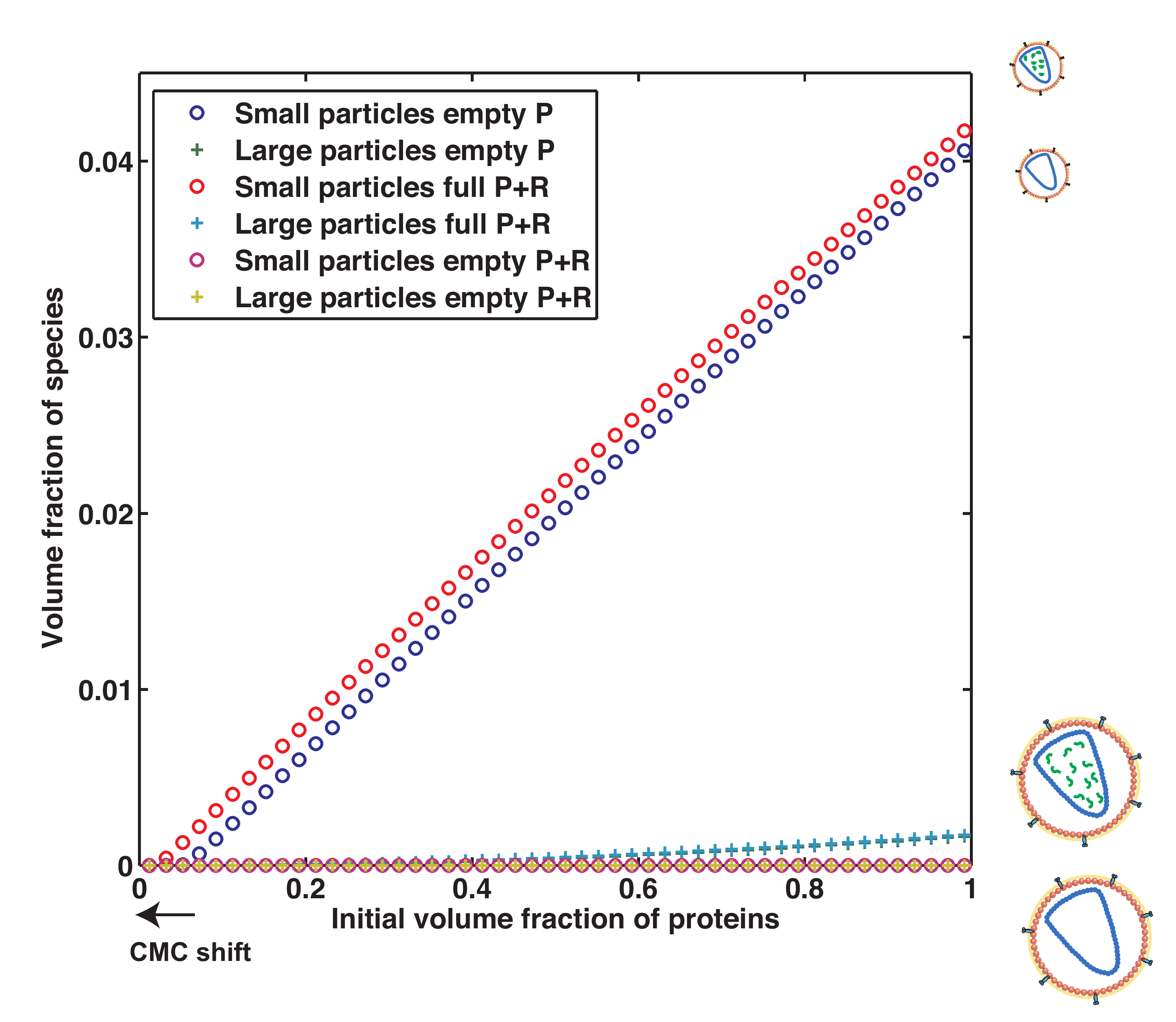}
\caption{Comparison of self-assembly of pure proteins and self-assembly of proteins in the presence of RNA. A shift of the onset of self-assembly by the presence of RNA is observed. This shift has been highlighted using a black arrow. Parameters are: $p_1=21,p_2=42,g_1^{(P)}=g_2^{(P)}=-2.5,g_1^{(P+R)}=g_2^{(P+R)}=-3.5,v0=2nm^3,\phi_rv_0=0.8$. }
\label{figure3}
\end{center}
\end{figure}

The second important feature of protein self-assembly in the presence of the genome is a shift of the CMC for particle assembly as compared to pure protein self-assembly. Indeed, the effective free energy gain per protein $g_{ieff}$ will contribute to the shift of  CMC, according to the previous estimation of  CMC in Eq. \ref{CMC}. This is clearly illustrated in the figure \ref{figure3} by comparing the two types of self-assembly. In this case, the self-assembly of proteins in the presence of RNA starts at lower protein concentration when compared to the self-assembly of pure proteins (cf figure \ref{figure3}). Note that for the parameters of figure 3, RNA uptake is assumed to reduce the free energy per protein ($\delta g_i \neq 0$), thereby reducing further $g_{ieff}$.

\section{Viral self-assembly in the presence of viral and cellular RNAs}
\subsection{Entropic selection of viral RNAs}
The uptake of viral genome during capsid self-assembly is made through both electrostatic and specific interactions. The former interaction is largely responsible for the linear relation between protein numbers and nucleotides observed in viruses databases. On the other hand, virologists have identified for many viral genome some specific sequence that have stronger affinities with viral proteins than electrostatic-based predictions\cite{aronoff1991,qu1997,comas2012}. This sequence is called a packaging signal (PSI or $\psi$). As a consequence, many viruses may contain both viral RNA bearing the $\psi$ sequence and non-viral cellular RNAs. This has been indeed observed for several viruses, and in particular for retroviruses like HIV-1 \cite{muriaux2001,rulli2007}. We anticipate in this case that the entropy of viral and cellular RNA will be of utmost relevance in determining the size distribution of particles and their RNA content. These phenomena can be described within the framework of micellization thermodynamics similarly to the discussion of previous sections.
 
 We consider in this section a mixture of proteins, monodisperse viral RNAs and cellular RNAs of respective initial concentrations $\phi_0,\phi_{rv}$ and $\phi_{rc}$. 
The main difference between two types of RNA within our simple model is their length: viral RNAs are usually longer than cellular RNAs \cite{yoffe2008}. Since each particle has the ability to contain both RNAs, we use for the particle with index $i$ a generalized linear relation between protein numbers $p_i$, viral RNAs $n_i$ and cellular RNAs $m_i$ such that
\begin{equation}
p_i=K_vn_i+K_cm_i
\end{equation}
In particular, the ratio $K_v/K_c$ scales like the ratio of RNA length. The equilibrium equations describing self-assembly are generalized from previous sections into
\begin{eqnarray}
\phi_0 & = & c_0+\sum_i p_ic_{p_i} \nonumber \\
\phi_{rv} & = & c_{rv}+\sum_in_ic_{p_i} \nonumber \\
\phi_{rc}& = & c_{rc}+\sum_im_ic_{p_i}\nonumber\\
c_{p_i} & = & c_0^{p_i}c_{rv}^{n_i}c_{rc}^{m_i}e^{-p_ig_i} \nonumber \\
\end{eqnarray}
where $c_{rv}$ and $c_{rc}$ are respectively the concentration of free viral and cellular RNAs.
These non-linear equations do not have general analytical solutions and may lead to a large variety of molecule partitioning among particles. Rather, it is possible to infer the influence of multiple RNA inside the particles by restricting the final products of self-assembly. In particular, imposing particles of same size but with different RNA content as the final product of self-assembly allows to address the question of preferential uptake of multiple RNAs as function the RNA partitioning. Therefore we restrict the analysis in this section to two final products of self-assembly: particles made of $p_1$ proteins and containing $n_1$ large viral RNAs and $m_1$ small cellular RNAs, and particles made of $p_2=p_1$ proteins and containing only $m_2$ small cellular RNAs and no large viral RNAs.
\begin{figure}[htbp]
\begin{center}
\includegraphics[scale=0.35]{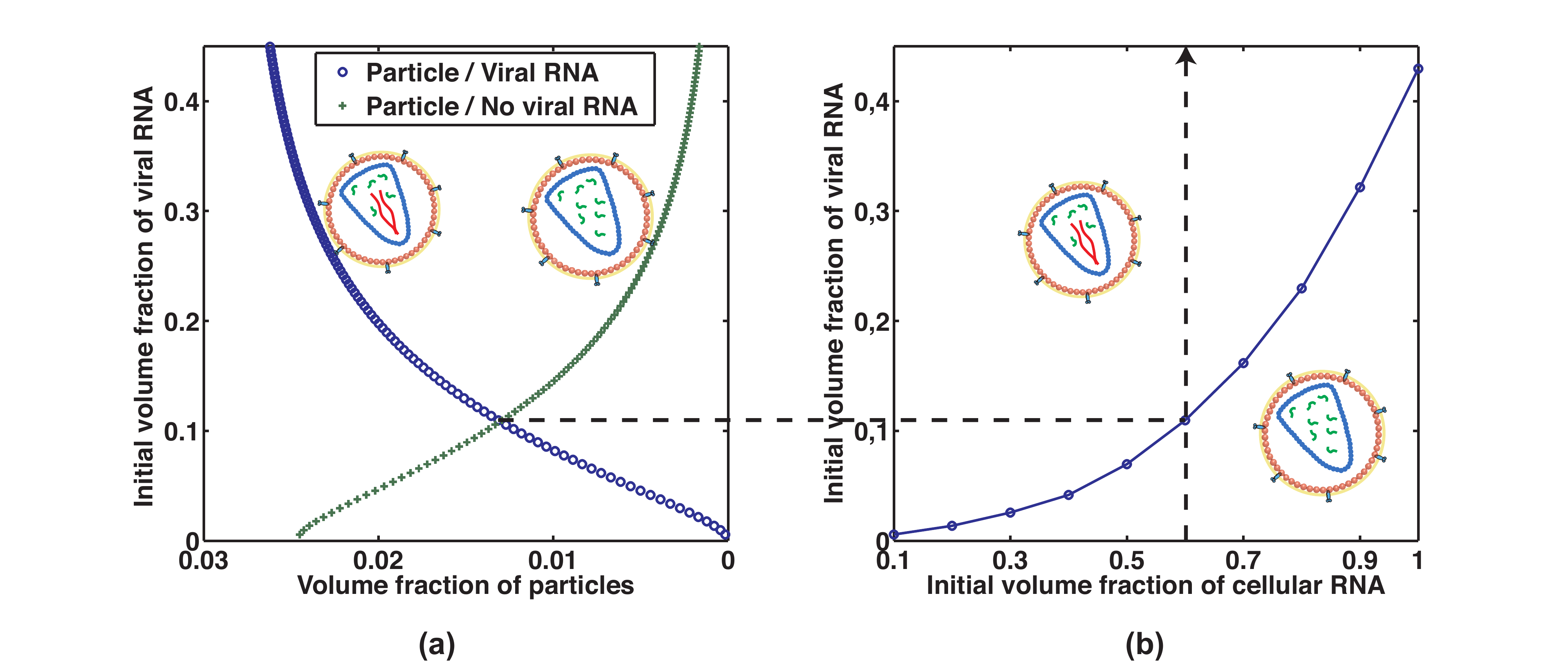}
\caption{Entropic selection of viral genome. \textit{(a)} Titration of viral RNA at fixed initial protein volume fraction $\phi_0v_0=0.8$ and fixed cellular RNA volume fraction $\phi_{rc}v_0=0.6$. Above some threshold concentration shown by the black dashed line, the particles containing the viral RNA are dominant. \textit{(b)} Corresponding phase diagram $\phi_{rv}/\phi_{rc}$ at fixed protein volume fraction. The line with circular symbols is the boundary where the volume fraction of particles containing viral and cellular RNAs equals the volume fraction of particles containing only cellular RNAs. The dashed arrow shows the line along which the graph of \textit{(a)} has been extracted. Other common parameters to \textit{(a)} and \textit{(b)} are: $p_1=p_2=21,n_1=2,m_1=10,n_2=0,m2=14,K_v=3,K_c=1.5,g_1=g_2=-2,v_0=2nm^3,\phi_0v_0=0.8$. }
\label{figure4}
\end{center}  
\end{figure}

The typical results in this case of viral RNA titration at fixed concentration of proteins and cellular RNAs is shown in figure \ref{figure4}a. In this case, there exists a threshold above which the uptake of viral RNA is systematically favorable. Interestingly, due to the length difference between RNAs, the threshold concentration is much smaller than the actual cellular RNA concentration. This can be understood qualitatively by the following entropic argument. The replacement of several small cellular RNAs by some longer viral RNAs in order to maintain the constant level of nucleotides required for a given capsid size allows to reduce effectively the number of small cellular RNAs per particle. Therefore a larger number of particles with reduced number of cellular RNAs can be made at constant cellular RNA concentration, and this entropically favorable.

Our results shows that the length difference between viral and cellular RNA is prone to favor viral RNA uptake based solely on entropic considerations. This is remarkable than without any sequence specificity, the spontaneous tendency of the viral self-assembly behavior is the uptake of
longer genome. The packaging signal $\psi$ adds another contribution to the preference of viral RNA\footnote{In the presence of a specific packaging signal, the free energy per protein $g_i$ is further reduced, therefore promoting particle formation incorporating this RNA, as compared to the case where no packaging signal is present.}, but it is not necessary to have a strong signal according to the previous entropic argument. Not surprinsingly, this entropic preference of large RNAs into particles disappears as the length difference between viral and cellular RNAs reduces (data not shown). 

\subsection{Entropic control of size distribution}
\begin{figure}[htbp]
\begin{center}
\includegraphics[scale=0.33]{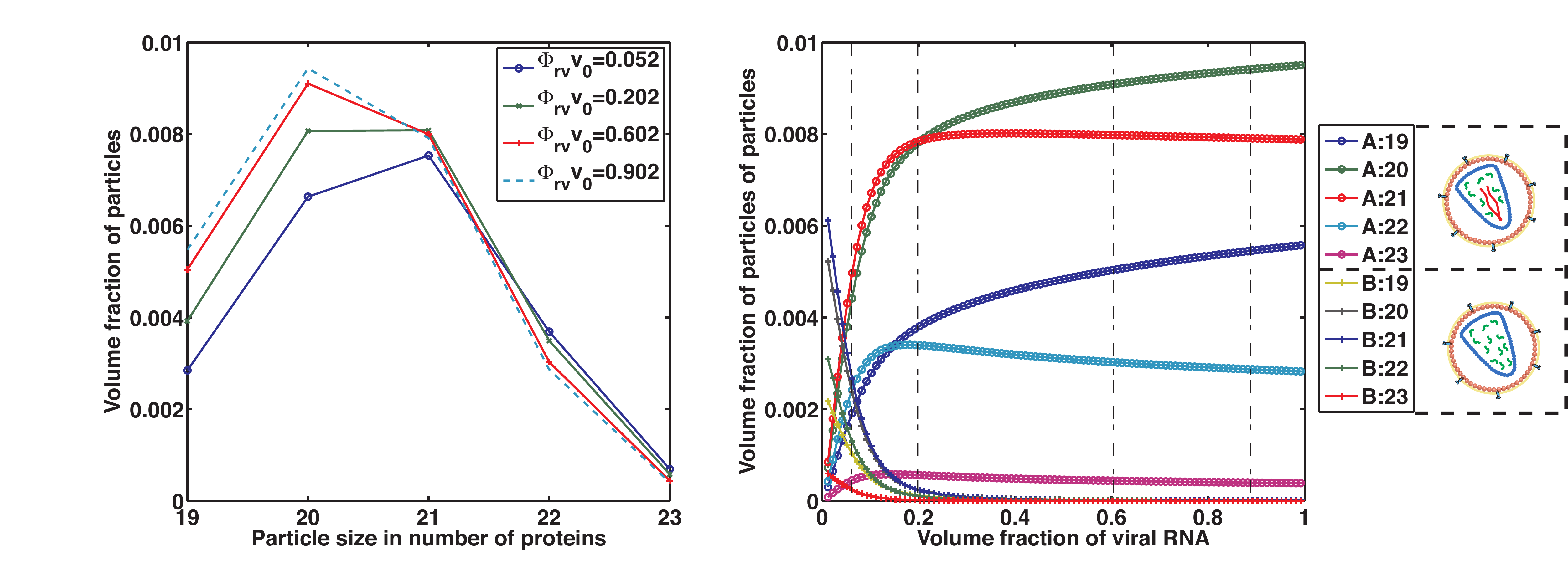}
\caption{Entropic control of size distribution. Five sizes of particles have been chosen for both A and B family. \textit{(a)} Evolution of viral particle size distribution (including A and B particles for each size) as the viral RNA concentration is increased. The population of particles shifts towards smaller sizes and the size polydispersity is reduced. \textit{(b)} Volume fraction of particles as function of their identity of particles as the viral RNA concentration is increased. The vertical dashed lines indicate the viral RNA volume fraction at which the size distribution of \textit{(a)} was extracted. Parameters are: $p_i=\{19,20,21,22,23\},n_A=2,m_A^{(i)}=\frac{p_i-K_+n_A}{K_-},m_B^{(i)}=\frac{p_i}{K_-},K_v=3,K_c=1.5,g_i=\{-1.92,-1.98,-2,-1.98,-1.92\},vo=2,\phi_0v_0=0.8,\phi_{rc}v_0=0.4$.}
\label{figure5}
\end{center}  
\end{figure}

By restricting the final product of self-assembly to a different set of particles, it is possible to go beyond simple bimodal products of self-assembly, and to describe the influence of the mixture of viral and cellular RNAs on the size distribution of viral particles. In order to investigate this case, we assume two families of particles: particles $A$ which have discrete sizes distributed evenly around a central value $p_0$ with a width $H\times\delta$, and particles $B$  which have the same discrete size distribution, but a different RNA content. A second index is introduced in order to label particles within each family $\{A^{(i)}\}$ and $\{B^{(i)}\}$. The particles $A^{(i)}$ contain $n_A^{(i)}$ viral RNAs and $m_A^{(i)}$ cellular RNAs, while the particles $B^{(i)}$ contain  $m_B^{(i)}$ cellular RNAs and no viral RNAs. Within this model, the number of particles of given size $p_i$ is now composed of particles $A$ and $B$. This particular classification of particles allow to solve numerically the equations for the size distributions while keeping informations on their identity (particles A or B). These equations are now written as
\begin{eqnarray}
\phi_0 & = & c_0+\sum_{i}p_i(c_{pA}^{(i)}+c_{pB}^{(i)}) \nonumber\\
\phi_{rv} & = & c_{rv}+\sum_{i}(n_A^{(i)}c_{pA}^{(i)}) \nonumber\\
\phi_{rc} & = & c_{rc}+\sum_{i}(m_A^{(i)}c_{pA}^{(i)}+m_B^{(i)}c_{pB}^{(i)}) \nonumber\\
c_{pA}^{(i)} & = & c_0^{p_i}c_{rc}^{m_A^{(i)}}c_{rv}^{n_A^{(i)}}e^{-p_ig_i}\nonumber\\
c_{pB}^{(i)} & = & c_0^{p_i}c_{rc}^{m_B^{(i)}}e^{-p_ig_i}\nonumber\\
\end{eqnarray}
A typical numerical solution for these equations is shown in figure \ref{figure5}a. 
The energetic parameters $g_i$ of the calculation were chosen in order to favor the central number of proteins $p_0$. As a consequence, the size distribution at low viral RNA concentration has a peak around $p_0$ of enthalpic origin. Interestingly, titration of viral RNA leads to a shift of the most favorable size towards smaller size. This is attributed to the entropic effects associated to the difference in RNA length between viral and cellular RNAs, similarly to the particular case of entropic selection discussed in the previous section. Moreover, since these entropic effects tend to favor smaller particles, the size polydispersity of this discrete set of particles is reduced. Notice that both effects of peak shift and polydispersity reduction reach a saturation, as it is explicitly seen in figure \ref{figure5}b. 
\section{Discussion}
We presented in the previous sections an analysis of the influence of viral genome in the self-assembly of proteins into capsid using the framework of micellization thermodynamics. Focusing on the specific effects associated to the entropy of partitioning all molecules (proteins and RNAs) among various particles, we identified several relevant features. The first one is that within a self-assembly scenario without inherent strong size selection of enthalpic origin, entropy will favor smaller particles. This is easily understood as more particles of smaller size can be made at constant number of proteins, and this is entropically favorable. This observation is of central importance since the presence of viral genome will essentially enhance this preference for smaller particles. 

More precisely the presence of monodisperse RNAs during viral self-assembly has been shown to enhance the preference for smaller particles, and to shift the CMC for viral self-assembly towards lower protein concentration. In the case where both viral and cellular RNAs are uptaken by viral particles, viral RNAs, which have been shown to be longer than most cellular RNAs \cite{yoffe2008}, are preferentially chosen for the self-assembly. Moreover, we showed that the size distribution of viral particles is shifted towards smaller sized particles and the size polydispersity is accordingly reduced.

Most of the findings described previously rely on the assumption of weak size selection of enthalpic origin. This assumption is certainly arguable in the case of most icosahedral virus, but it is likely to be realistic in the case of retroviruses like HIV-1. Indeed, the size distribution of HIV-1 has been shown to be quite large, reflecting the absence of a strong size-selection mechanism, whatever its precise origin \cite{moskalenko2012}. As a consequence, we might expect that some of the results of the present work are applicable to HIV-1. We were recently able to address this question experimentally using viral particles produced within cells, and by quantifying their size distribution thanks to Atomic Force Microscopy imaging \cite{moskalenko2012}. Remarkably, we found that viral particles grown in the presence of viral genome were statistically smaller than particles grown in its absence, and that the size polydispersity was also reduced, in qualitative agreement with the prediction of our models. Similarly, evidence of the entropic selection of large genome was observed in studies quantifying the RNA amount within HIV-1 particles \cite{muriaux2001,rulli2007}: in the absence of viral genome, a few number of large RNAs were observed within viruses (typically one or two). 

Interestingly recent observations on members of the Paramyxoviruses family, like the Newcastle Disease Virus (NDV) are also qualitatively explained by the entropic features highlighted in our work \cite{goff2012}: indeed, it was observed in this case that a majority of infectious VLPsare small and contain a single genome, while a minority are large and contain multiple genomes. The qualitative observation of effects predicted by the entropy of partitioning of molecules during self-assembly shows therefore unambiguously that the entropy contributes to the control of viral particle size distribution.

The authors would like to thank the \textit{Fondation Simone et Cino Del Duca} from the \textit{Institut de France} for initial funding that allowed to launch this project. This work was also partially supported thanks to CNRS program entitled ``PIR: Interface physique, biologie et chimie''.

\bibliographystyle{unsrt}
\bibliography{bibHDR}

\begin{thebibliography}{10}

\bibitem{lyonnais2012}
S.~Lyonnais, R.J. Gorelick, F.~Heniche-Boukhalfa, S.~Bouaziz, V.~Parissi, J.F.
  Mouscadet, T.~Restle, J.M. Gatell, E.~Le Cam, and G.~Mirambeau.
\newblock A protein ballet around the viral genome orchestrated by hiv-1
  reverse transcriptase leads to an architectural switch: From
  nucleocapsid-condensed {RNA} to vpr-bridged {DNA}.
\newblock {\em Virus Res.}, 2012.

\bibitem{zandi2004}
R.~Zandi, D.~Reguera, R.F. Bruinsma, W.M. Gelbart, and J.~Rudnick.
\newblock Origin of the icosahedral symmetry in viruses.
\newblock {\em Proc. Nat. Acad. Sci.}, {\bf 101}:15556--15560, 2004.

\bibitem{siber2009}
A.~\v{S}iber and A.~Majdandzic.
\newblock Spontaneous curvature as a regulator of the size of virus capsids.
\newblock {\em Phys. Rev. E}, {\bf 80}:021910, 2009.

\bibitem{zandi2009}
R.~Zandi and P.~Van der Schoot.
\newblock Size regulation of ss-{RNA} viruses.
\newblock {\em Biophys. J.}, {\bf 96}:9--20, 2009.

\bibitem{belyi2006}
V.~A. Belyi and M.~Muthukumar.
\newblock Electrostatics origin of the genome packing in viruses.
\newblock {\em Proc. Nat. Acad. Sci.}, {\bf 103}:17174--17178, 2006.

\bibitem{ting2011}
C.L. Ting, J.~Wu, and Z.G. Wang.
\newblock Thermodynamic basis for the genome to capsid charge relationship in
  viral encapsidation.
\newblock {\em Proc. Nat. Acad. Sci.}, {\bf 108}:16986--16991, 2011.

\bibitem{hu2008}
Y.~Hu, R.~Zandi, A.~Anavitarte, C.M. Knobler, and W.M. Gelbart.
\newblock Packaging of a polymer by a viral capsid: the interplay between
  polymer length and capsid size.
\newblock {\em Biophys. J.}, {\bf 94}:1428--1436, 2008.

\bibitem{aronoff1991}
R.~Aronoff and M.~Linial.
\newblock Specificity of retroviral {RNA} packaging.
\newblock {\em J. Virol.}, {\bf 65}:71--80, 1991.

\bibitem{qu1997}
F.~Qu and T.J. Morris.
\newblock Encapsidation of turnip crinkle virus is defined by a specific
  packaging signal and {RNA} size.
\newblock {\em J. Virol.}, {\bf 71}:1428--1435, 1997.

\bibitem{comas2012}
M.~Comas-Garcia, R.D. Cadena-Nava, A.L.N. Rao, C.M. Knobler, and W.M. Gelbart.
\newblock In vitro quantification of the relative packaging efficiencies of
  single-stranded{RNA} molecules by viral capsid protein.
\newblock {\em J. Virol.}, {\bf 86}:12271--12282, 2012.

\bibitem{muriaux2001}
D.~Muriaux, J.~Mirro, D.~Harvin, and A.~Rein.
\newblock {RNA} is a structural element in retrovirus particles.
\newblock {\em Proc. Nat. Acad. Sci.}, {\bf 98}:5246--5251, 2001.

\bibitem{rulli2007}
S.~Rulli, C.S. Hibbert, J.~Mirro, T.~Pederson, S.~Biswal, and A.~Rein.
\newblock Selective and nonselective packaging of cellular {RNA}s in retrovirus
  particles.
\newblock {\em J. Virol.}, {\bf 81}:6623--6631, 2007.

\bibitem{yoffe2008}
A.M. Yoffe, P.~Prinsen, A.~Gopal, C.M. Knobler, W.M. Gelbart, and A.~Ben-Shaul.
\newblock Predicting the sizes of large {RNA} molecules.
\newblock {\em Proc. Nat. Acad. Sci.}, {\bf 105}:16153--16158, 2008.

\bibitem{moskalenko2012}
C.~Faivre-Moskalenko, A.~Thomas, K.~Tartour, J.~Bernaud, M.~Iazykov, Y.~Beck,
  J.~Danial, M.~Lourdin, D.~Muriaux, and M.~Castelnovo.
\newblock {RNA} control of hiv-1 particle size polydispersity.
\newblock {\em submitted}, 2012.

\bibitem{goff2012}
P.H. Goff, Q.~Gao, and P.~Palese.
\newblock A majority of infectious newcastle disease virus particles contain a
  single genome, while a minority contain multiple genomes.
\newblock {\em J. Virol.}, {\bf 86}:10852--10856, 2012.

\end{thebibliography}
\end{document}